\documentclass{aa}

\usepackage{natbib}
\usepackage{graphicx}
\usepackage{txfonts}
\usepackage{hyperref}

\usepackage{rotating}
\usepackage{xcolor}
\usepackage{amsmath}
\usepackage{nccmath}
\usepackage{subfig}
\usepackage{fixltx2e}
\usepackage[utf8]{inputenc}
\newcommand{\oii}{$[$\ion{O}{II}$]$}

\begin{document} 
\title{Towards sub-kpc scale kinematics of molecular and ionized gas of star-forming galaxies at $z\sim1$}

\author{M. Girard\inst{1}
\and
M. Dessauges-Zavadsky\inst{1}
\and
F. Combes\inst{2}
\and
J. Chisholm\inst{3}
\and
V. Patr\'icio\inst{4}
\and
J. Richard\inst{5}
\and
D. Schaerer\inst{1,6}
}

\institute{Observatoire de Genève, Université de Genève,
51 Ch. des Maillettes, 1290 Sauverny, Switzerland\\
\email{mgirard@swin.edu.au}
\and
Observatoire de Paris, LERMA, Collège de France, CNRS, PSL Univ., Sorbonne University, UPMC, Paris, France
\and
University of California--Santa Cruz, 1156 High Street, Santa Cruz, CA, 95064, USA
\and 
DARK, Niels Bohr Institute, University of Copenhagen, Lyngbyvej 2, 2100 Copenhagen, Denmark
\and
Univ Lyon, Univ Lyon1, Ens de Lyon, CNRS, Centre de Recherche Astrophysique de Lyon UMR5574, F-69230, Saint-Genis-Laval, France
\and      
CNRS, IRAP, 14 Avenue E. Belin, 31400 Toulouse, France
}

\date{Received - ; accepted - }

\abstract{ We compare the molecular and ionized gas kinematics of two strongly lensed galaxies at $z\sim1$ that lie on the main sequence at this redshift, based on observations from ALMA and MUSE, respectively. We derive the CO and \oii \, rotation curves and dispersion profiles of these two galaxies. We find a difference between the observed molecular and ionized gas rotation curves for one of the two galaxies, the Cosmic Snake, for which we obtain a spatial resolution of few hundred parsecs along the major axis. The rotation curve of the molecular gas is steeper than the rotation curve of the ionized gas. In the second galaxy, A521, the molecular and ionized gas rotation curves are consistent, but the spatial resolution is only of few kpc on the major axis. Using simulations, we investigate the effect of the thickness of the gas disk  and effective radius on the observed rotation curves and find that a more extended and thicker disk smooths the curve. We also find that the presence of a strongly inclined (>70$^\circ$) thick disk (>1 kpc) can smooth the rotation curve because it degrades the spatial resolution along the line of sight. By building a model using a stellar disk and two gas disks, we reproduce the rotation curves of the Cosmic Snake with a molecular gas disk that is more massive and more radially and vertically concentrated than the ionized gas disk. Finally, we also obtain an intrinsic velocity dispersion in the Cosmic Snake of $18.5\pm7$  $\rm km \, s^{-1}$   and $19.5\pm6$  $\rm km \, s^{-1}$ for the molecular and ionized gas, respectively, which is consistent with a molecular disk with a smaller and thinner disk. For A521, the intrinsic velocity dispersion values are $11\pm8$  $\rm km \, s^{-1}$ and $54\pm11$  $\rm km \, s^{-1}$, with a higher value for the ionized gas. This could indicate that the ionized gas disk is thicker and more turbulent in this galaxy. These results highlight the diversity of the kinematics of galaxies at $z\sim1$ and the different spatial distribution of the molecular and ionized gas disks. It suggests the presence of thick ionized gas disks at this epoch and that the formation of the molecular gas is limited to the  midplane  and center of the galaxy in some objects.
}

\keywords{galaxies: high-redshift -- galaxies: kinematics and dynamics --  gravitational lensing: strong}

\maketitle

\section{Introduction}

The kinematics of star-forming galaxies in the local Universe provide insight into their physical conditions \citep[e.g.,][]{vanderKruit1978,Sofue2001,vanderKruit2011,Glazebrook2013, BlandHawthorn2016}. Their rotation curves can particularly give information on the different structures present in the galaxies, such as a nuclear bar \citep[e.g.,][]{Sorai2000}, but also on the fraction and distribution of baryonic and dark matter  \citep[e.g.,][]{vandeHulst1957,Schmidt1957,Ford1971, Bosma1979, Rubin1983,Carignan2006,deBlok2008}.

The kinematics of star-forming galaxies can be measured using the hydrogen line at 21 cm, CO lines, nebular emission lines and stellar absorption lines. These lines trace the neutral hydrogen gas (\ion{H}{I}), molecular gas (H$_2$), ionized gas and the stars, respectively. Previous studies have found similar \ion{H}{I}, CO, and H$\alpha$ rotation curves in the disks of local spiral galaxies \citep[][]{Sofue1996,Sofue1999}. However, small discrepancies between H$\alpha$ and CO can occur in the inner regions due to the extinction and contamination from the bulge affecting H$\alpha$ \citep{Sofue2001}. \citet{Frank2016} observed small discrepancies between the CO and \ion{H}{I} kinematics in the inner part of some galaxies, and attributed the discripancies to the presence of a bar that affects only the CO emission. The \ion{H}{I} is also sometimes missing in the inner regions. \citet{Richards2016} found that the \ion{H}{I} rotation curve rises more slowly than the H$\alpha$ rotation curve in most of the galaxies when the neutral gas disk is more extended than the ionized gas disk. 
Comparing the molecular and ionized gas rotation curves of 17 rotation-dominated galaxies from the Extragalactic Database for Galaxy Evolution (EDGE) and Calar Alto Legacy Integral Field Area (CALIFA) surveys, respectively, \citet{Levy2018} obtained that 75\% of their galaxies have a smaller rotation velocity for the ionized gas than for the molecular gas. They argued that it could be due to a vertical rotation in a thick disk or diffuse extraplanar ionized gas.

\begin{table*}[tb]
\caption[]{Galaxy properties}
\label{table1}
\centering                          
\begin{tabular}{l c c c c c c c c}        
\hline\hline                
Name     &  $\alpha$ & $\delta$    & PSF$\rm_{MUSE}$ & PSF$\rm_{ALMA}$ & $z$   & $\rm M_\star$   & SFR\\
         &  (J2000.0) & (J2000.0)  &  [$''$]         &  [$''$]         &       & $\rm [M_\odot]$ & $\rm [M_\odot \, yr^{-1}]$\\
\hline                     
\noalign{\smallskip}
Cosmic Snake & 12:06:11 & -08:48:05 & 0.51 & $0.22\times0.18$  &  1.03620   &  $(4.0\pm0.5) \times 10^{10}$ & $30.0\pm10$ \\
A521 & 04:54:06 & -10:13:23 & 0.57 & $0.20\times0.17$  &  1.04356  &    $(5.4\pm1.3)\times 10^{10}$ & $15\pm4$ \\
\noalign{\smallskip}
\hline    
\end{tabular}
\tablefoot{The stellar mass, $\rm M_\star$, and star formation rate, SFR, have been obtained by SED fitting \citep{Cava2018, Patricio2018} and are corrected for the magnification. The PSF$\rm_{MUSE}$ indicates the seeing during the observations and PSF$\rm_{ALMA}$ indicates the synthesized beam size.}
\end{table*}


At high redshift, the easiest way to study the kinematics is to observe strong emission lines tracing the ionized gas \citep[e.g.,][]{ForsterSchreiber2009,Wisnioski2015,Stott2016, Swinbank2017, Turner2017, Girard2018a}. The \ion{H}{I}, CO and stellar absorption lines are more difficult to detect in  distant galaxies, but with recent advances in instrumentation, it is now possible to measure the CO kinematics \citep[e.g.,][]{Rivera2018}, and the stellar kinematics \citep[e.g.,][]{Guerou2017}. For example, \citet{Barro2017} presented the molecular gas kinematics of the center (<1 kpc) of a galaxy at $z=2.3$. The velocity map revealed a velocity gradient consistent with a rotating disk. They found $\upsilon_{rot}/\sigma_0\sim2.5$, where $\upsilon_{rot}$ is the maximum rotation velocity and $\sigma_0$ is the intrinsic velocity dispersion. This means that this galaxy is dominated by rotation ($\upsilon_{rot}/\sigma_0>1$) within the central 1 kpc. To get the kinematics at higher redshift ($z>4$), some studies have also observed the [\ion{C}{II}] \, emission line \citep[e.g.,][]{Matthee2017,Smit2018}. For example, \citet{Smit2018} found two galaxies that are potentially rotating at $z\sim6.8$.


Observing these new tracers at high redshift allows us to compare them to the ionized gas kinematics, similar to what has been done in the local Universe.
Using deep observations from NOEMA and LBT, \citet{Ubler2018} found a good agreement between the molecular and ionized gas kinematics for a massive galaxy at $z=1.4$. 
Similarly, \citet[][]{Guerou2017} compared the ionized gas and stellar kinematics of 17 galaxies at $0.2<z<0.8$ and also found consistent kinematics. 
\citet{Swinbank2011} used CO observations to identify a rotation-dominated galaxy, but \citet{Olivares2016} obtained a very irregular H$\alpha$ kinematics with no sign of ordered motion for the same object. It is therefore still unclear if the ionized gas traces well the kinematics in these high-redshift galaxies and why discrepancies between different kinematic tracers are observed in some galaxies.

The intrinsic velocity dispersion has also been measured from both the molecular and ionized gas in nearby galaxies and at high redshift. For example, \citet{Cortese2017} compared CO and H$\alpha$ velocity dispersions of massive star-forming galaxies at $z \sim 0.2$. They found that the H$\alpha$ velocity dispersion is higher in most of the galaxies ($ \sim$20-40 $\rm km \, s^{-1}$) than the CO velocity dispersion ($\sim$10-20 $\rm km \, s^{-1}$), with values comparable to what is observed in the local Universe. Additionally, \citet{Ubler2018} obtained similar values ($\sim$15-30 $\rm km \, s^{-1}$) of the velocity dispersion in CO and H$\alpha$ for their massive galaxy at $z\sim1.4$. \citet{Ubler2019} also found that both molecular gas and ionized gas dispersion increases at high redshift. According to this study, the ionized gas dispersion is higher than the molecular gas dispersion by $\rm \sim$10-15$\rm \, km \, s^{-1}$ on average at all redshifts.

Recent studies have used gravitational lensing to study the kinematics at higher spatial resolution and  to analyse low-mass galaxies at high redshift \citep[e.g.,][]{Jones2010, Livermore2015, Leethochawalit2016, Mason2017, Girard2018a, Patricio2018, Girard2018b}. For example, \citet{Leethochawalit2016} reached a resolution of <500 pc for 15 low-mass galaxies (log($M_\star/M_\odot)\sim9.0$-9.6) at $z\sim2$. They found that their velocity maps are highly disturbed.
\citet{Patricio2018} analysed eight highly magnified galaxies for which they also obtained a spatial resolution of a few hundred parsecs. They found for these galaxies with log($M_\star/M_\odot)\sim10.0$-10.9 a smooth rotating disk with a ratio of $\upsilon_{rot}/\sigma$ between 2 and 10, typical for galaxies at $0.6<z<1.5$.

In this work, we present the kinematic analysis of two strongly lensed galaxies from the \citet{Patricio2018} sample, the Cosmic Snake and A521, at $z=1.036$ and $z=1.044$, located behind the clusters MACSJ1206.2-0847 and Abell 0521, respectively. A detailed study of the stellar clumps and the molecular clouds of the Cosmic Snake has already been performed, using HST \citep{Cava2018} and ALMA observations (Dessauges-Zavadsky et al. accepted). The ionized gas kinematics of both galaxies have also been studied by \citet{Patricio2018} using MUSE observations. By combining the CO(4-3) and \oii \, observations, we can compare the molecular and ionized gas kinematics of these two galaxies at high spatial resolution (a few hundred parsecs on average).

The paper is organized as follows: Section \ref{section2} describes the previous studies of these two galaxies. In Sect. \ref{section3}, we explain the observations, data reduction and gravitational lens modeling. The comparison of the molecular and ionized gas kinematics is described in Sect. \ref{section4}. Sections \ref{section5} and \ref{sect_analysis} present the analysis and discussion of our results. We finally present our conclusions in Sect. \ref{section6}.

In this paper, we adopt a $\Lambda$-CDM cosmology with $\rm H_0 = 70  \, km \, s^{-1} \, Mpc^{-1}$, $\Omega_M = 0.3$, and $\Omega_\Lambda= 0.7$. We adopt a \citet{Salpeter1955} initial mass function (IMF).

%
\begin{sidewaysfigure*}
\centering
\includegraphics[width=\columnwidth]{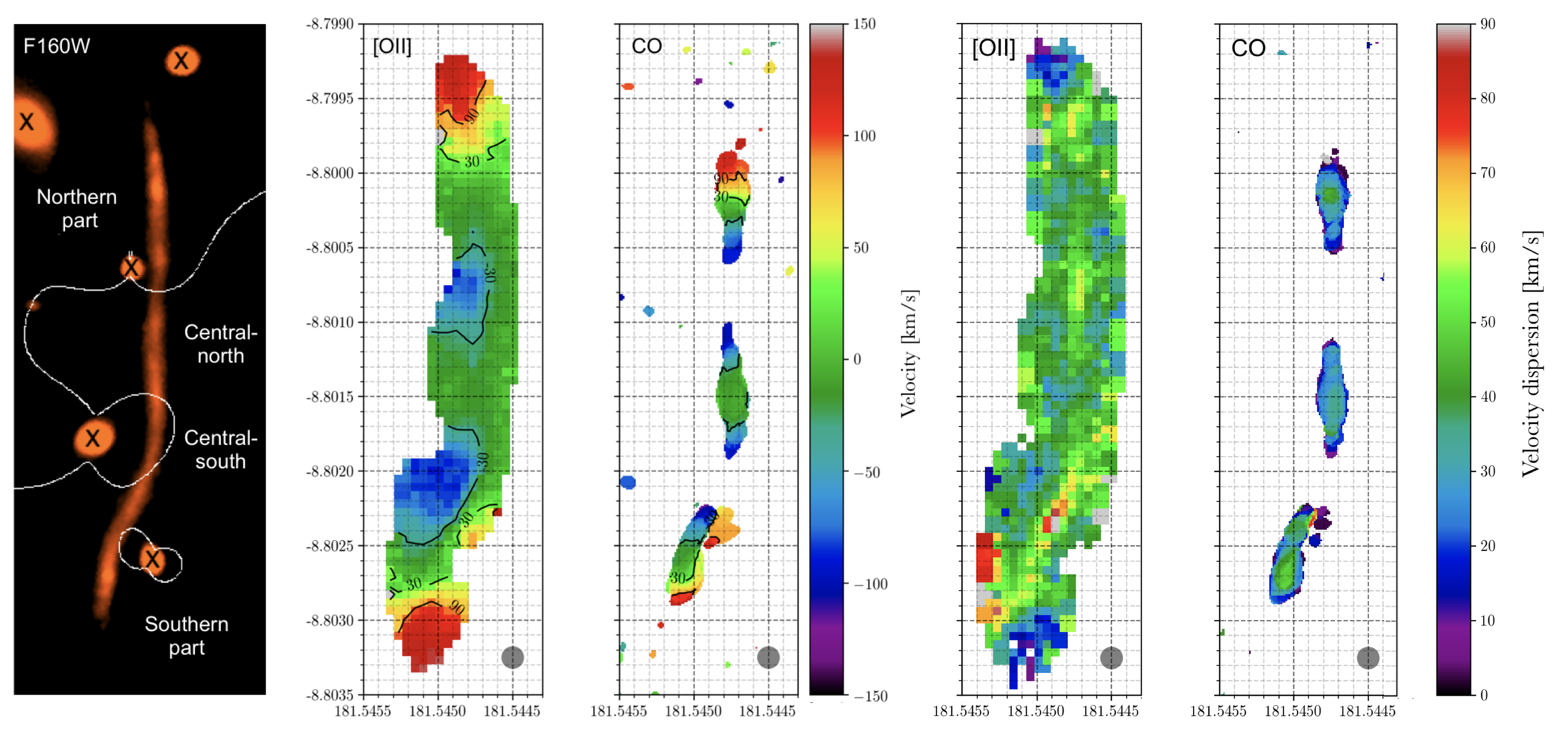}
\caption{HST/WFC3 F160W near-infrared image (left panel), \oii \, (middle left panel) and CO (middle panel) emission line velocity maps, and \oii \, (middle right panel) and CO (right panel) velocity dispersion maps of the Cosmic Snake in the image plane. The PSF  of $0.51''$ of the kinematic maps is shown as a grey circle at the bottom right. The black contours indicate velocities of 30 and 90 $\rm km\,s^{-1}$. The critical line from the lensing model is shown in white on the left panel. The HST image and kinematic maps are at the same scale. The velocity dispersion maps are corrected for the instrumental broadening, but uncorrected for beam-smearing.}
\label{vel_snake}
\end{sidewaysfigure*}

\begin{figure*}
\centering
\subfloat{\includegraphics[scale=0.31]{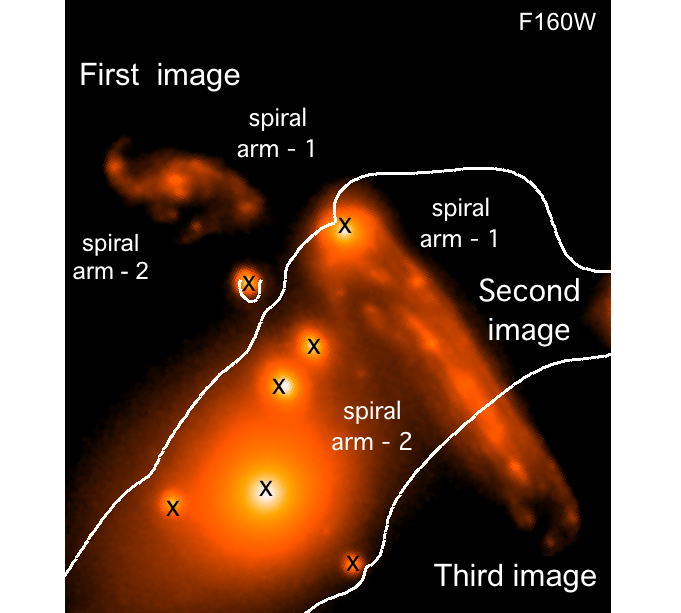}}\\
\subfloat{\includegraphics[scale=0.685]{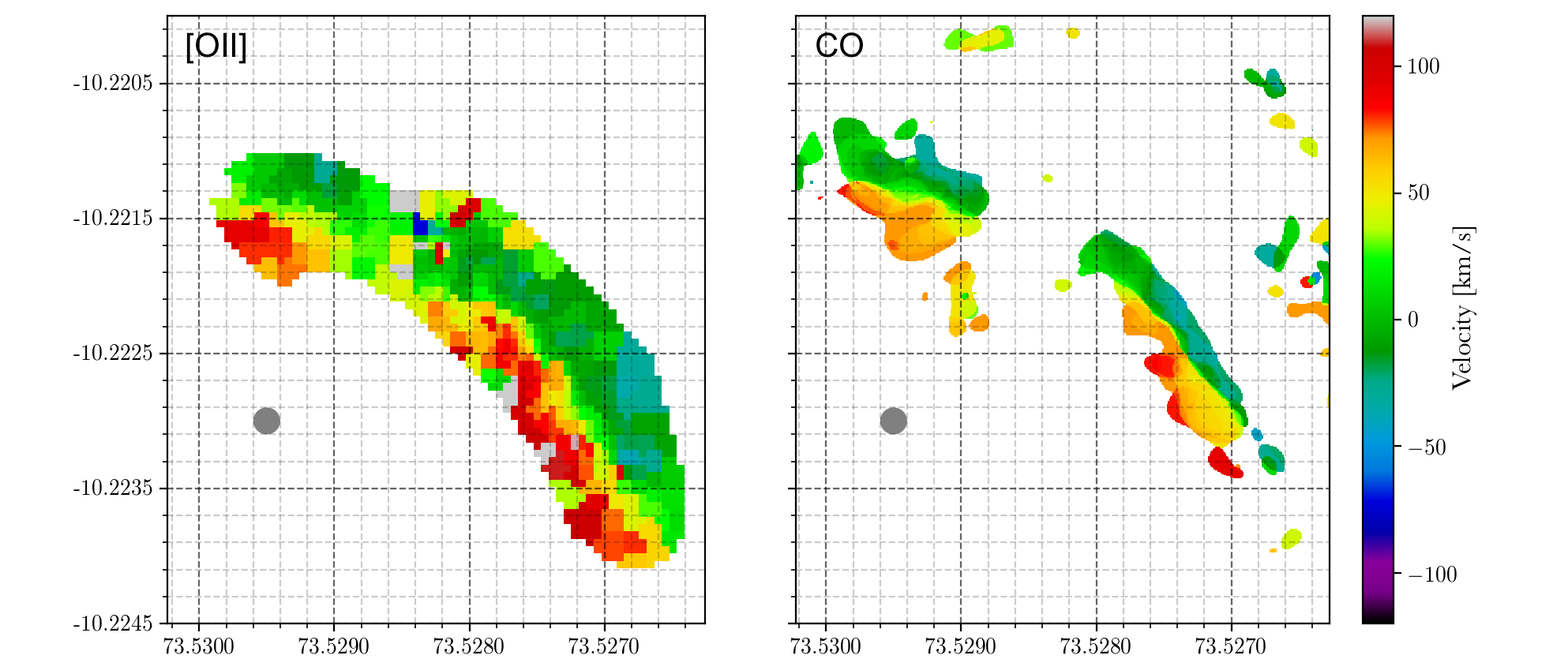}}\\
\subfloat{\includegraphics[scale=0.685]{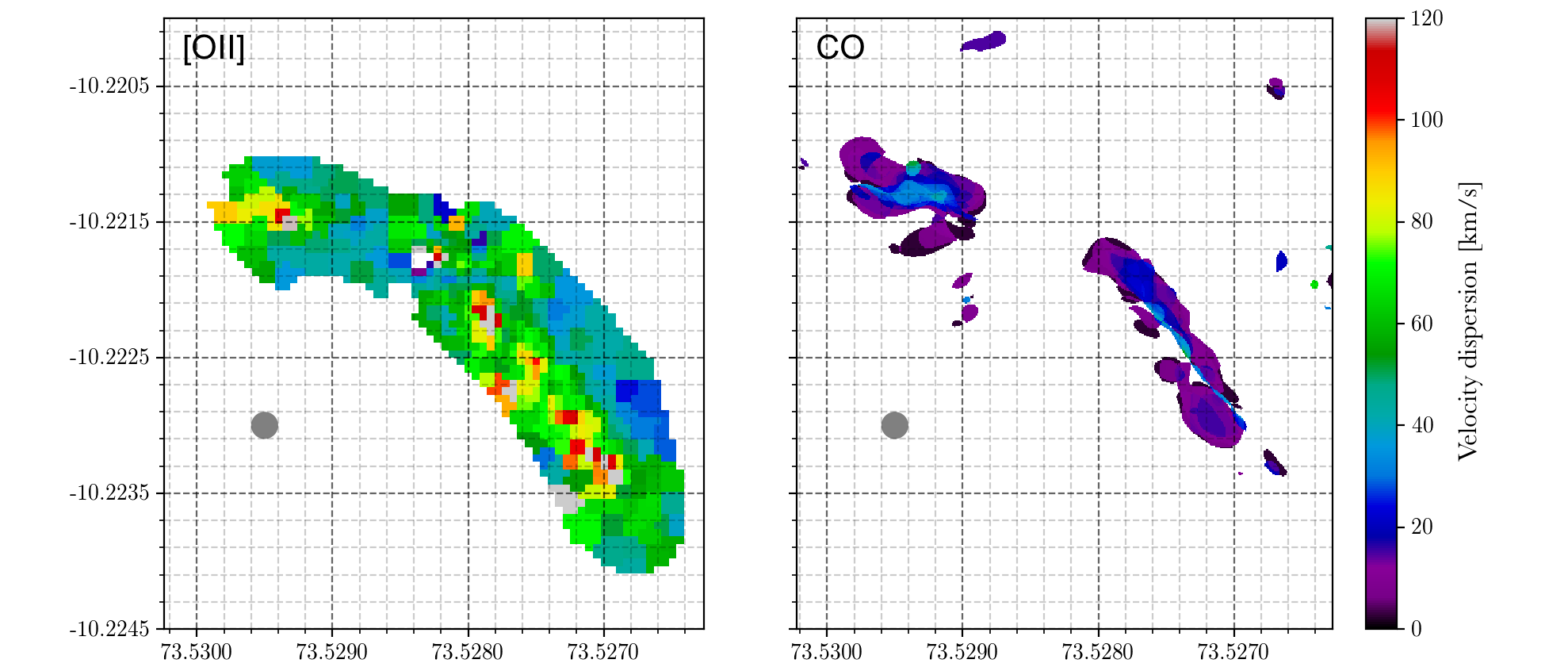}}
\caption{HST/WFC3 F160W near-infrared image (top panel), \oii \, emission line velocity map (middle left), CO line velocity map (middle right), \oii \, emission line velocity dispersion map (bottom left) and CO line velocity dispersion map (bottom right)  of A521 in the image plane. The PSF of $0.69''$ of the kinematic maps is shown as a grey circle at the bottom left. Both velocity dispersion maps are corrected for the instrumental broadening, but uncorrected for beam-smearing. The critical line from the lensing model is shown in white on the top panel. The HST image and kinematic maps are at the same scale. }
\label{A521_maps}
\end{figure*}

\section{The Cosmic Snake and A521}
\label{section2}

The Cosmic Snake and A521 are two strongly lensed galaxies located behind the clusters MACSJ1206.2-0847 and Abell 0521. The foreground cluster bends and magnifies the light from the background galaxy, forming multiple images stretched into arcs on the sky of the same region within the background galaxy with magnification factors that vary from a few to hundreds. 

The Cosmic Snake shows a four-folded multiple image of the same galaxy aligned along an arc. The northern and southern parts present two counter-images of a visible fraction larger than 50\% of the source galaxy (see Fig. \ref{vel_snake}). The central part presents two counter-images, central-north and central-south, both showing a fraction of less than 20\% of the source galaxy. From the CO kinematic maps (see Fig. \ref{vel_snake}), we clearly see the northern, central (that includes central-north and central-south), and the southern parts. 

A521 is a typical disk galaxy that shows clear spiral arms \citep{Richard2010,Patricio2018}. This galaxy is lensed into three counter-images of the same galaxy. The first counter-image is located in the north east (see Fig. \ref{A521_maps}) and shows a spiral galaxy barely distorted, with two clear spiral arms. The two other counter-images are more distorted and elongated due to higher magnifications and are located to the west. The second counter-image in the north east region shows the same first spiral arm, and in the south region a second spiral arm. The third image is smaller and represents only the last couple of spaxels in the south west of the kinematic maps. In contrast to the Cosmic Snake where we detect the CO emission only in the center of the galaxy, here the CO traces the spiral arms. Again, from the CO kinematic maps (right panels of Fig. \ref{A521_maps}), we clearly distinguish the east part (which includes the first counter-image) and the west part (which includes the second and third counter-images).

The main properties of the Cosmic Snake and A521 including their stellar mass, $\rm M_\star$, star formation rate, SFR, and redshift $z$, are presented in Table \ref{table1}. The two galaxies show several star-forming clumps (see \citet{Cava2018} for the Cosmic Snake) and lie on the main sequence of star-forming galaxies at this redshift (see Fig. 3 in \citet{Patricio2018}). 
They are also included in the sample studied by \citet{Patricio2018}, who measured inclinations of 70$^\circ$ and 72$^\circ$ from kinematic models, respectively.
Using a model that takes account of the lensing and  beam-smearing effects, they found $\upsilon_{rot}/\sigma_0$ ratios of $5.1\pm3.6$ and $2.1\pm0.8$ for the Cosmic Snake and A521, respectively, which are typical at $z\sim1$ \citep[e.g.,][]{Wisnioski2015}.  This suggests that these two galaxies are disks dominated by rotation.

\section{Observations, data reduction and analysis}
\label{section3}
 
\subsection{\oii \, observations with MUSE/VLT}

The observations, data reduction and emission line measurements are described in \citet{Patricio2018}. The observations were seeing-limited for the Cosmic Snake while A521 was observed with adaptive optics. The PSF obtained during the observations were $0.51''$  and $0.57''$, respectively (see Table \ref{table1}). 

They derived the velocity field and velocity dispersion maps of the Cosmic Snake and A521 by fitting the \oii$\lambda3726,3729 \AA$ emission lines with Gaussian profiles in the image plane using the CAMEL code \citep{Epinat2010}. They used a Voronoi binning when the signal to noise ratio was lower than five in a spaxel in A521 \citep{Cappellari2003}. The A521 maps have been slightly smoothed during this process. Therefore, the final PSF of the Cosmic Snake and A521 are $\sim0.51''$ and $\sim0.69''$, respectively. The spectral resolution of MUSE is $\rm \sim50 \, km \, s^{-1}$.

\subsection{CO observations with ALMA}

Observations of the Cosmic Snake were carried out in Cycle 3 (project 2013.1.01330.S)  and were performed in the extended C38-5 configuration with the maximum baseline of 1.6 km and 38 of the 12 meters antennae. The total on-source integration time was of 52.3 minutes in band 6. We observed the CO(4-3) emission line at the observed frequency of 226.44 GHz which corresponds to a redshift of $z=1.036$. A521 was observed in Cycle 4 (project 2016.1.00643.S) using the C40-6 configuration with the maximum baseline of 3.1 km and 41 12 meters antennae. Similarly, the observations were carried out in band 6 to detect the CO(4-3) emission line at an observed frequency of 225.66 GHz, corresponding to a redshift of $z=1.043$, with an on-source time of 89.0 minutes. The spectral resolution was tuned to 7.8125 MHz ($\rm \sim 10.3\, km \, s^{-1}$) for both galaxies.

The data reduction was performed with the standard automated reduction procedure from the pipeline of the Common Astronomy Software Application (CASA) package \citep{McMullin2007}. To image the CO(4-3) line, we used the Briggs weighting and the robust factor of 0.5, which gives a good compromise between the resolution and sensitivity. We cleaned all channels interactively, with the \emph{clean} routine in CASA, until convergence. To perform the cleaning, we used a custom mask for which we made sure to include the CO emission of all the channels. We then applied the primary beam correction. The final synthesized beam size is $0.22'' \times 0.18'' $ with a position angle of +85$^\circ$ for the Cosmic Snake and $0.20'' \times 0.17'' $ at $-71^\circ$ for A521. We reach an rms of 0.29 mJy\,beam$^{-1}$ and 0.34 mJy\,beam$^{-1}$ per 7.8125 MHz channel for the Cosmic Snake and A521, respectively.

Both the Cosmic Snake and A521 datacubes were smoothed with the \emph{imsmooth} routine in CASA to a spatial resolution of $0.51''$ and $0.69''$, respectively, to get the same final spatial resolution as obtained in the MUSE data. We finally use the \emph{immoments} routine in CASA to obtain the CO(4-3) moment maps, with a 3$\sigma$ threshold parameter above the noise level applied when producing the first (velocity) and second (velocity dispersion) moment maps.

\begin{figure*}[t]
\centering
\subfloat{\includegraphics[scale=0.43]{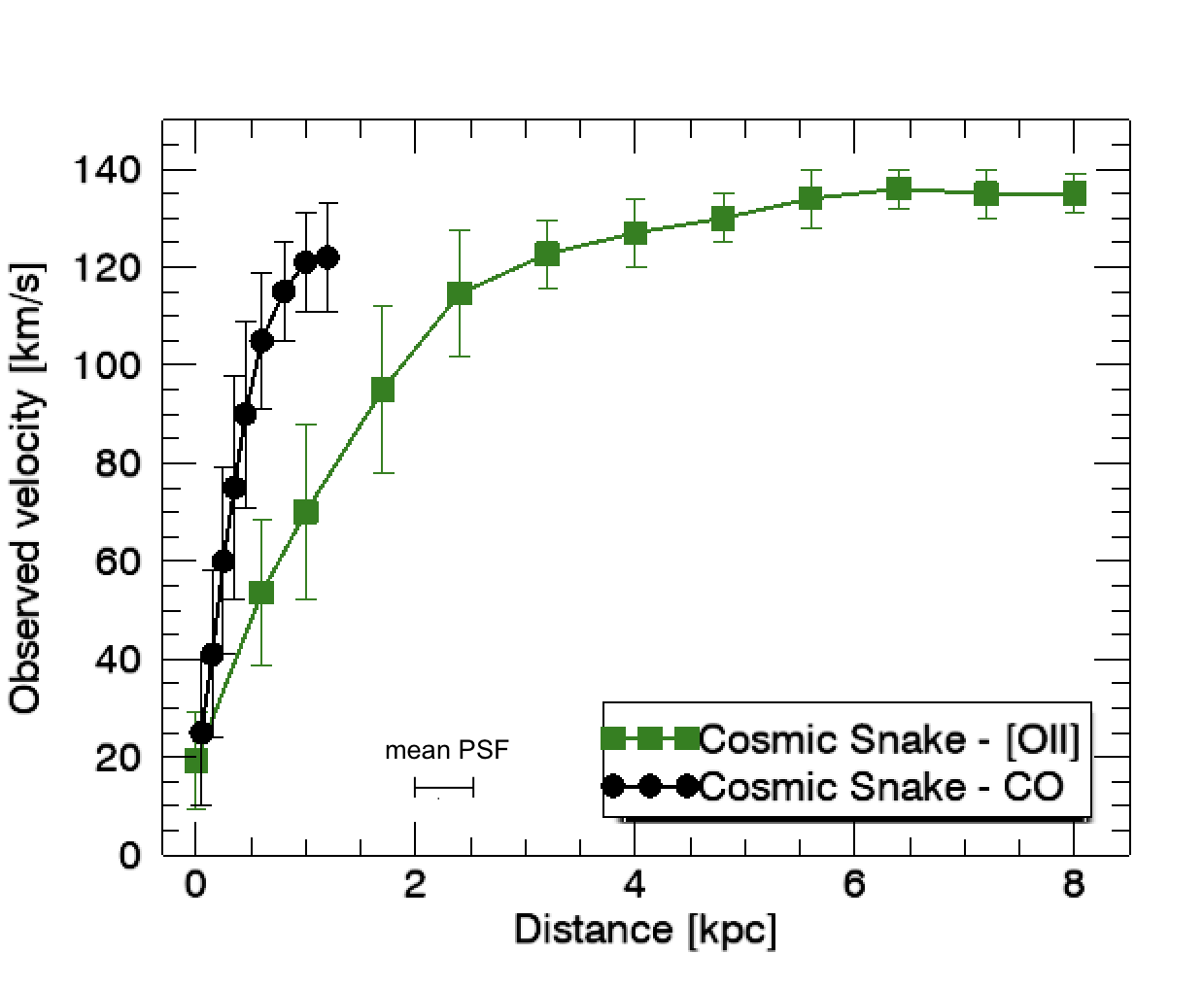}}
\subfloat{\includegraphics[scale=0.43]{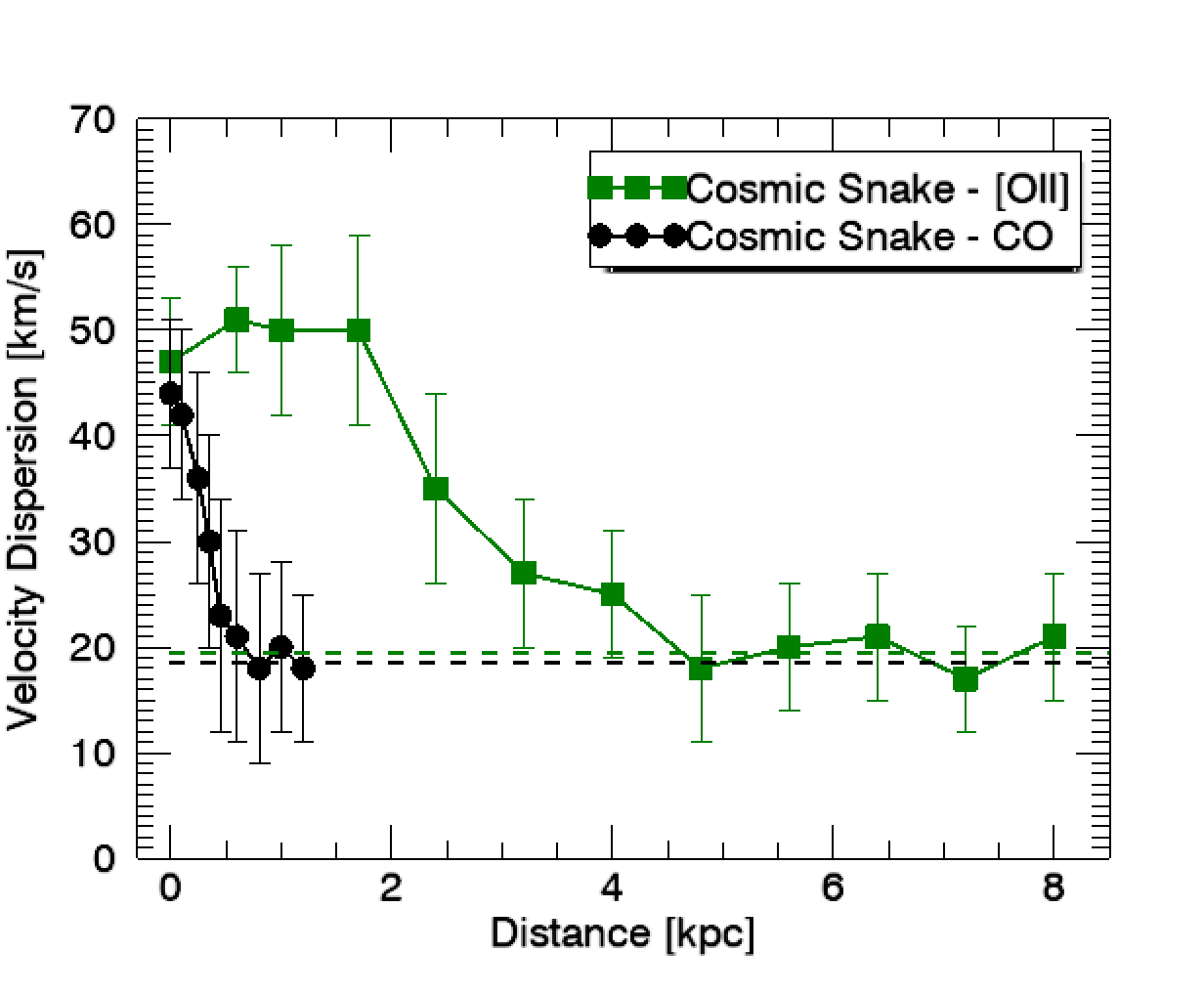}}
\caption{ Observed velocity (left panel) and velocity dispersion profiles (right panel) of the Cosmic Snake from the \oii \, emission line (in green) and CO(4-3) line (in black). The solid line is the average rotation curve of the two counter-images. The curves are extracted from the observed data in the image plane, uncorrected for beam smearing or inclination. Both CO and \oii \, datacubes have been convolved to the same spatial resolution of $0.51''$ ($\sim 500$ pc on average). The two  dashed lines indicate the intrinsic velocity dispersion of $19.5\pm 6 $  $\rm km \, s^{-1}$   and $18.5\pm 7 $  $\rm km \, s^{-1}$   measured from the \oii \, (in green) and CO (in black) velocity dispersion maps at large radii ($R>R_e$ when possible),  respectively.  }
\label{profil_snake}
\end{figure*}
\begin{figure*}[t]
\centering
\subfloat{\includegraphics[scale=0.43]{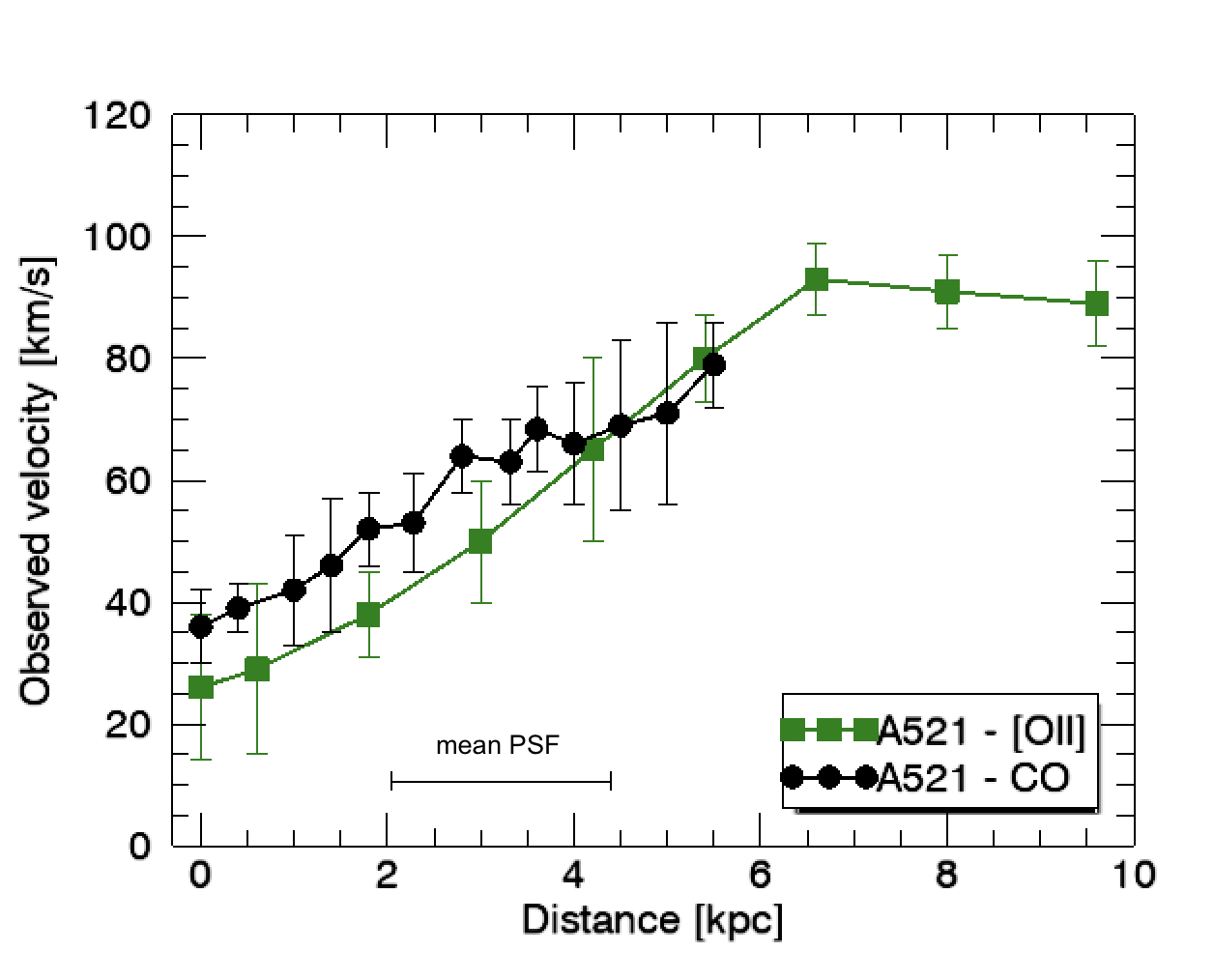}}
\subfloat{\includegraphics[scale=0.43]{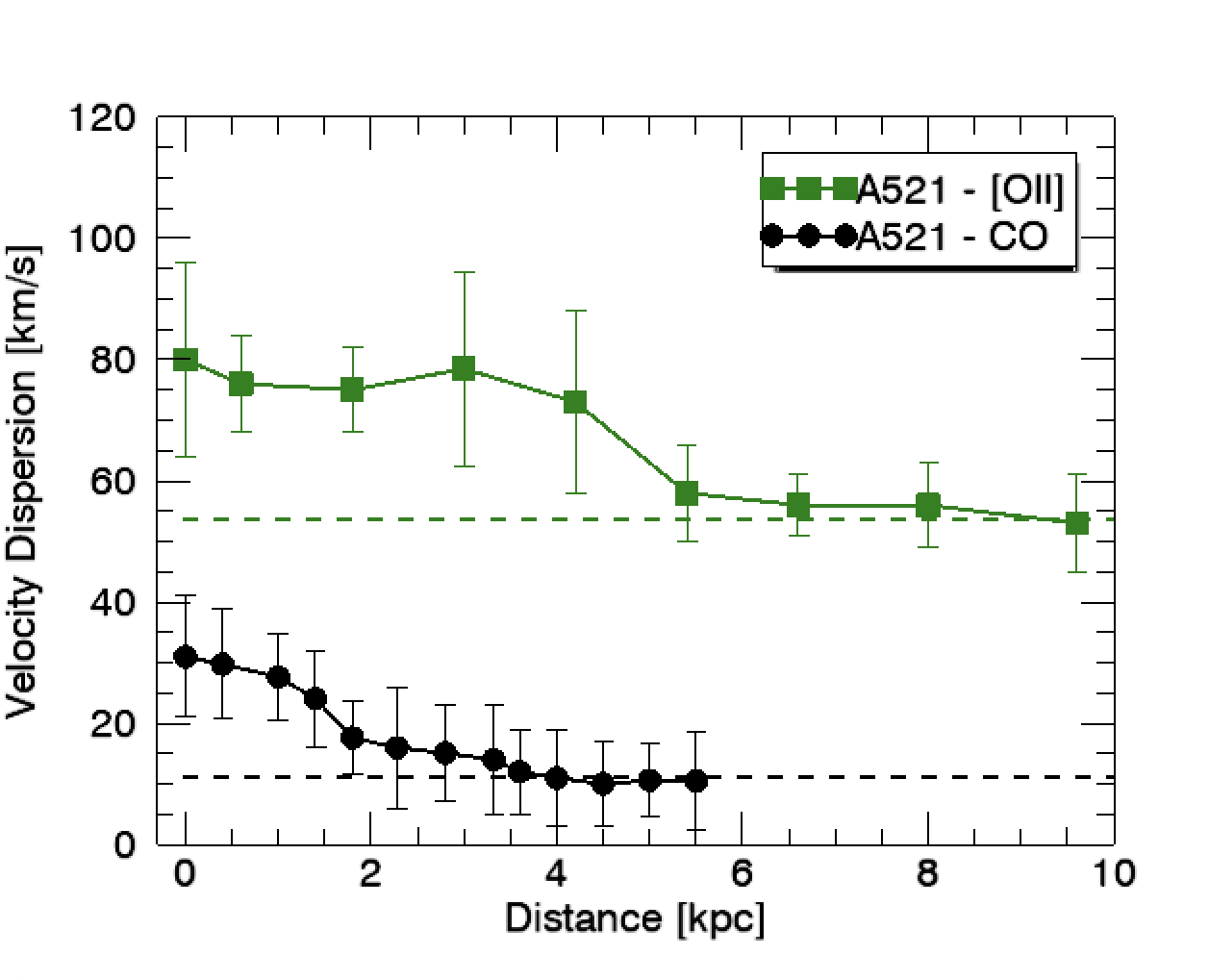}}
\caption{ Observed velocity (left panel) and velocity dispersion profiles (right panel) of A521 from the \oii \, emission line (in green) and CO(4-3) line (in black). The solid line is the average rotation curve of the two counter-images. The curves are extracted  from the observed data in the image plane, uncorrected for beam smearing or inclination. Both CO and \oii \, datacubes have been convolved to the same spatial resolution of $0.69''$ ($\sim 2-3$ kpc on average). The two  dashed lines indicate the intrinsic velocity dispersion of $54 \pm 11 $  $\rm km \, s^{-1}$   and $11 \pm 8 $  $\rm km \, s^{-1}$   measured from the \oii \, (in green) and CO (in black) velocity dispersion maps at large radii ($R>R_e$ when possible), respectively. }
\label{profil_a521}
\end{figure*}

\subsection{Gravitational lens modeling}
\label{lens}

The two galaxies presented in this work are highly magnified and have multiple images, therefore we need to use a lensing model to recover the physical galactocentric radius corrected for the lensing magnification. We use the lensing models that were presented in detail in \citet{Cava2018} and \citet{Richard2010} for the Cosmic Snake and A521, respectively, which were based on multiple images identified in HST imaging. Lenstool was used to constrain the mass distribution and to optimize the models for these two galaxies using their multiple images. Uncertainties on the magnification vary between 10-20\%. The lens model achieves an rms precision of $0.15''$ and $0.23''$ between the predicted and observed locations of the strong lensing constraints as measured in the image plane of the Cosmic Snake and A521 arcs, respectively.


\section{Comparison of the molecular and ionized gas kinematics}
\label{section4}

Figures \ref{vel_snake} and \ref{A521_maps} show the observed velocity and velocity dispersion maps of the Cosmic Snake and A521 in the image plane at a spatial resolution of $\sim 0.51 ''$ and $\sim 0.69 ''$, respectively, with the HST/WFC3 F160W images (program \#15435). The \oii \, and CO(4-3) kinematic maps have been measured in the galaxy rest-frame using the redshifts indicated in Table \ref{table1}. The velocity dispersion maps have been corrected for their corresponding instrumental broadening, $\sigma_{instr}$, using the following relation:

\begin{ceqn}
\begin{align}
\label{LSF_correction}
\sigma = \sqrt{\sigma_{obs}^2 - \sigma_{instr}^2}.
\end{align}
\end{ceqn}
 
To measure the rotation curves and velocity dispersion profiles, we construct concentric annuli in the source plane. The width of these annuli has been chosen to be slightly larger than the physical spaxel size. We then project them in the image plane with spaxels of the same size as the ALMA and MUSE data. Hence, we get the physical distance to the center in the source plane for every spaxel. We define the major axis as the line following the direction of the steepest velocity gradient using the maximum and minimum values in each annulus on the velocity map. These major axis correspond well to the ones derived in \citet{Patricio2018}. The rotation curves and velocity dispersion profiles have been measured directly from the velocity and velocity dispersion maps in the image plane by taking an aperture corresponding to the PSF along the major axis. This method has been verified using a simulated galaxy in the source plane and projected in the image plane. We accurately recovered the kinematics (with an error less than 10\%).  To measure the final rotation curves, we take only the mean values of the positive side of the rotation curves of two counter-images for each galaxy that show the highest visible fraction of the source galaxy: the north and south parts of the Cosmic Snake, and the north east part (first image) and the east image of the west part (second image) of A521. We do not include negative velocities because less than 20\% of the negative velocities from the background galaxies are lensed (see Sect. \ref{section2}).

Figures \ref{profil_snake} and \ref{profil_a521} present the observed rotation curves, uncorrected for the inclination, and velocity dispersion profiles, uncorrected for the beam-smearing, of the CO and \oii \, emission lines of the Cosmic Snake and A521, respectively. The solid line represents the average values obtained from the two counter-images in both cases. The error bars have been determined from the velocity measurement on the two counter-images for the two galaxies. The profiles of the Cosmic Snake and A521 have an average spatial resolution of few hundred parsecs ($\sim500$ pc) and few kilo-parsecs ($\sim2-3$ kpc), respectively. Since the major axis of A521 is perpendicular to the stretching direction, we do not get as high of spatial resolution as for the Cosmic Snake.

In Fig. \ref{profil_snake}, we see a clear difference between the molecular and ionized gas rotation curves of the Cosmic Snake. The molecular gas curve is steeper and  reaches the maximum rotation velocity at a smaller radius from the center than the ionized gas curve. On the other hand, A521 does not show a difference between the CO and \oii \, rotation curves (Fig. \ref{profil_a521}).

We determine the intrinsic velocity dispersion by directly measuring the velocity dispersion maps along the major axis in the outer regions (at radii larger than the effective radius, $R_e$, when possible) where the rotation curve is flat and beam-smearing is negligible \citep[e.g.,][]{ForsterSchreiber2009, Girard2018a}. 
We obtain a similar intrinsic velocity dispersion of $18.5 \pm 7 \, $  $\rm km \, s^{-1}$   and $19.5 \pm 6 $  $\rm km \, s^{-1}$  for the molecular and ionized gas, respectively, for the Cosmic Snake. For A521, we find different values of $11 \pm 8 $  $\rm km \, s^{-1}$   and $54 \pm 11 $  $\rm km \, s^{-1}$, respectively. \citet{Patricio2018} used a different method to estimate the intrinsic velocity dispersion where they corrected in quadrature the beam smearing determined from the modeled velocity dispersion. They obtained values of $44 \pm 31 $  $\rm km \, s^{-1}$  and $63 \pm 23 $  $\rm km \, s^{-1}$ for the ionized gas intrinsic velocity dispersion in the Cosmic Snake and A521, respectively.

\section{Modeled gas rotation curves}
\label{section5}

\begin{figure}[t]
\centering
\includegraphics[scale=0.58]{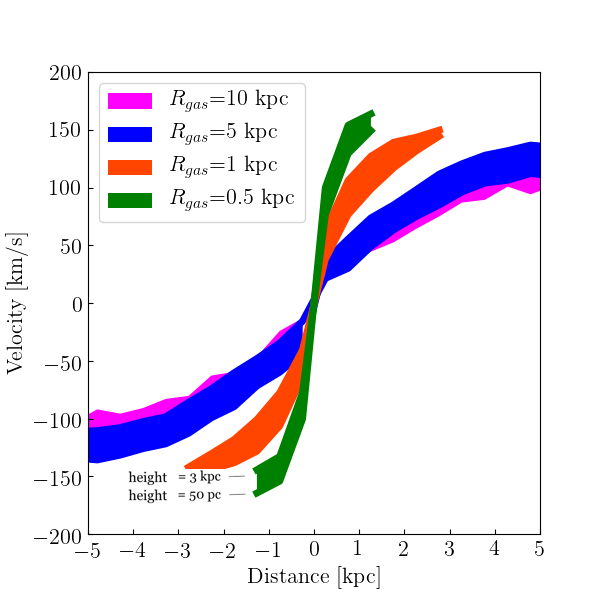}
\caption{Effect of the gas disk effective radius, $R_{gas}$, and height, $H_{gas}$, on the rotation curve. The values of the gas disk effective radius are 0.5 kpc (green), 1 kpc (orange), 5 kpc (blue) and 10 kpc (pink). The upper and lower boundaries (taking the absolute values) of each coloured curve represent a disk height of 50 pc and 3 kpc, respectively. The other parameters are fixed to $M_\star=3\times10^{10} \, M_\odot$, $R_\star=4.0$ kpc, $H_\star=0.5$ kpc, $f_{gas}=0.2$, $i=70^\circ$, and Q=1.}
\label{varie_rgas}
\end{figure}

We focus here on how the geometrical effects, such as the effective radius and thickness of the gas disk, can influence the shape of the rotation curve, and how this can cause differences in rotation curves of distinct gas tracers.

\subsection{Model}
\label{sect_rot_curve_model}

In order to build a realistic and self-consistent rotation curve model, due to a disk of stars and gas, and possibly a bulge or dark matter component, we choose to describe the various mass components by analytical density-potential pairs, and then perform a numerical simulation of gas particles in the resulting potential. This allows us to have more flexibility to study the various gas tracers, that can then sample the same rotation curve with different weights. For the stellar and gaseous disks, we choose to represent them by Miyamoto-Nagai disks of particles \citep{Miyamoto1975}, with masses, radial scales and heights corresponding to the observations. For the simulation, we typically distribute one million particles according to the analytical distributions of the gas disk components, with their relative effective radii and scale heights. The gas particles are distributed in nearly circular orbits, in equilibrium with the total potential, with a velocity distribution corresponding to the Toomre Q-parameter. The ratio between tangential and radial velocity dispersion was taken from the epicyclic theory \cite[e.g.,][]{Toomre1964}.

To reconstruct a rotation curve from the particle positions and velocities, we assume an inclination and then produce a 2D velocity map by taking the average velocity of the particles along the line of sight in each spatial element. The velocity map can be adjusted to the same average physical spatial resolution as the observations (in kpc).  Afterwards, we can extract the rotation curve in a similar way as an observed velocity map by measuring the values along the major axis.

We first investigate the general effect of the effective radius and scale height of a gas disk, $R_{gas}$ and $H_{gas}$, on the rotation curve for a model including only a stellar disk and a single gas disk. We neglect the dark matter since we look only at the baryons in the center of a typical relatively massive galaxy. We also neglect the bulge because its effect on the molecular and ionized gas kinematics are negligible. This model uses 6 fixed parameters for the stellar and gas disks: the stellar mass, $M_\star$, the stellar disk effective radius, $R_\star$, the stellar disk scale height, $H_\star$, the gas mass,  $M_{gas}$, (or the gas fraction, $f_{gas}$, where $f_{gas}= M_{gas}$/($M_{gas}$+$M_\star$)),  the Toomre stability parameter, $Q$, and the inclination, $i$. For the stellar mass and gas fraction, we use values similar to the molecular gas disks of the Cosmic Snake and A521 to represent a typical galaxy at high redshift with $M_\star = 3 \times 10^{10} \, M_\odot$, $f_{gas}=0.2$, and an inclination of $i=70^\circ$. For the stellar disk, we adopt values established in the literature for nearby galaxies  with $R_\star=4.0$ kpc and  $H_\star=0.5$ kpc \citep[e.g.,][]{vanderKruit2011,Ma2017}. Finally, we fix the Toomre Q-parameter to a value of  1, which represents a quasi-stable disk for a thin gas disk \citep[e.g.][]{Genzel2011}.

Fig. \ref{varie_rgas} shows the rotation curves obtained for an effective radius of the gas disk of 0.5 kpc (green), 1 kpc (orange), 5 kpc (blue), and 10 kpc (magenta). The upper and lower boundaries (taking absolute values) of the plotted rotation curves for the four effective radii correspond to a gas disk height of 50 pc and 3 kpc, respectively. A disk height of 50 pc is typically observed for a molecular gas disk in the local Universe \citep[e.g.,][]{Glazebrook2013}. We adopt a maximum value of 3 kpc for the disk height, since the average and maximum disk height values reported for high-redshift galaxies are $\sim0.63$ kpc and $\sim2.5$ kpc, respectively \citep[][]{Elmegreen2017}.

The effective radius strongly affects the shape of the rotation curve. The rotation curve of a disk with a smaller radius reaches the maximal rotation velocity at a smaller radius than a more extended disk, meaning that the smaller the effective radius, the steeper the rotation curve.
We see that the thickness of the gas disk also affects the rotation curve. A thicker disk smooths the rotation curve, such that the curve becomes less steep with increasing thickness. The effect of the thickness is less pronounced than the effect of the effective radius.

\subsection{Effect of the thickness and inclination on the spatial resolution}
\label{sect_res_effect}

\begin{figure}[t]
\centering
\includegraphics[scale=0.58]{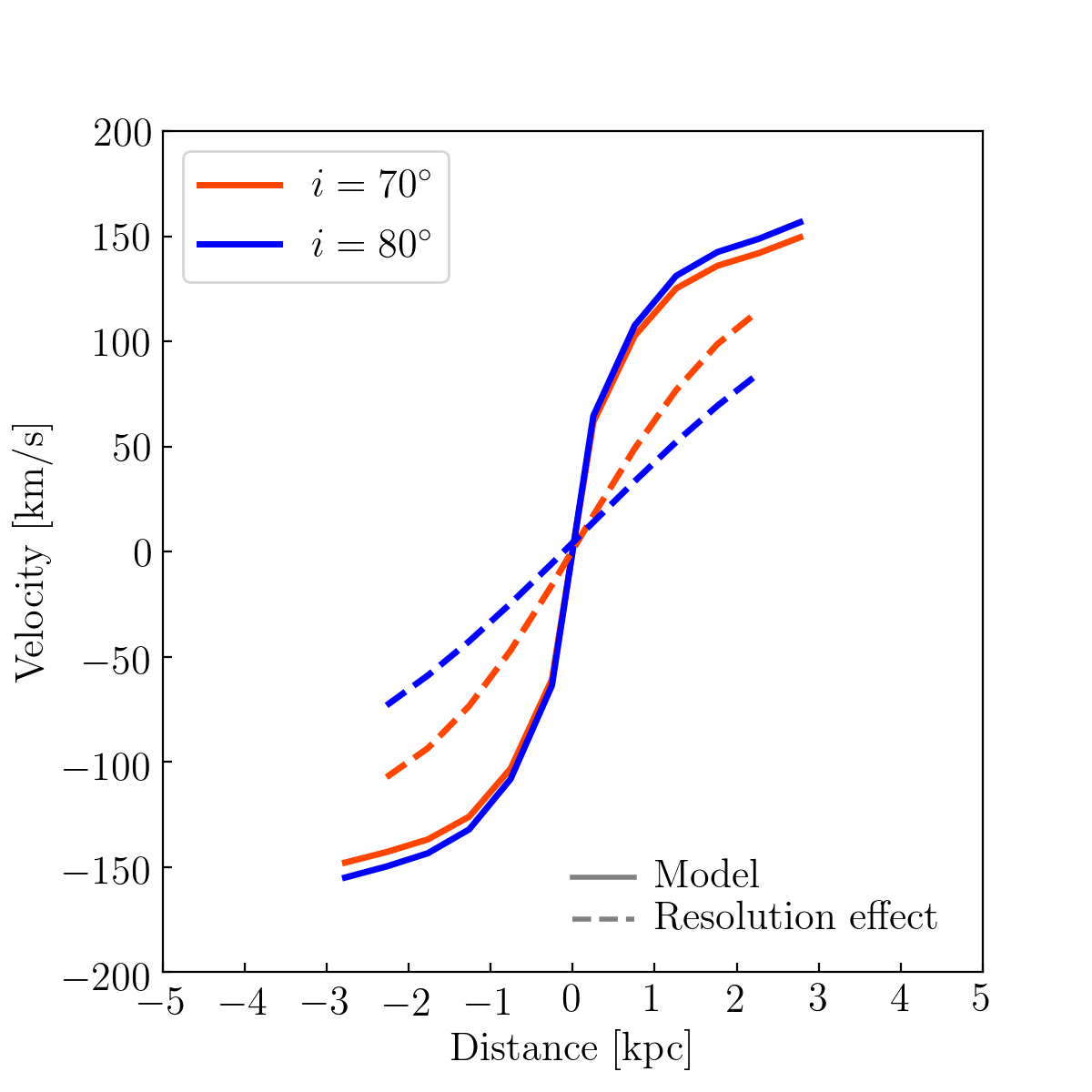}
\caption{Effect of the inclination on the rotation curve due to the spatial resolution. The solid curves represent the rotation curves simulated using the same  parameters as in Fig. \ref{varie_rgas}, with $R_{gas}=1$ kpc and $H_{gas}=1$ kpc for an inclination of $i=70^\circ$ (orange) and $i=80^\circ$ (blue). The dashed lines present the two models smoothed to the spatial resolution determined by Eq. \ref{resolution} with their corresponding inclinations and $H_{gas}=1$ kpc. }
\label{variePSF_i}
\end{figure}

The presence of a thick disk when a galaxy is strongly inclined limits the spatial resolution we can get on the observed rotation curve. The highest spatial resolution that can be obtained for an inclined thick disk can be described by:

\begin{ceqn}
\begin{align}
\label{resolution}
H \times tan( i ),
\end{align}
\end{ceqn}
where $H$ is the height of the disk, and $i$ is the inclination of the galaxy. This means that a galaxy with a thick disk and a high inclination could have a lower spatial resolution than the expected resolution from the observations. We investigate the effect of the spatial resolution on the rotation curve using the model described in Sect. \ref{sect_rot_curve_model}, with $R_{gas}$=1 kpc and $H_{gas}$=1 kpc. We convolve the 2D velocity maps of this model for an inclination of $i=70^\circ$ and $i=80^\circ$ to get their velocity maps at the spatial resolution determined by Eq. \ref{resolution} with their corresponding inclination and gas height. We then extract the respective rotation curves. Fig. \ref{variePSF_i} shows the simulated rotation curves for an inclination of $i=70^\circ$ (orange) and $i=80^\circ$ (blue) together with the rotation curves smoothed due to the lower spatial resolution. We see that the spatial resolution effect is much stronger for an inclination of $i=80^\circ$. Therefore, a high inclination also strongly smooths the rotation curve when a thick disk is present ($>~$1kpc). 

\section{Analysis of two lensed $z\sim1$ galaxies}
\label{sect_analysis}
In this section, we investigate if the difference observed between the rotation curves and velocity dispersion values of the molecular and ionized gas can be explained by geometrical effects in the Cosmic Snake and A521. 

\subsection{The Cosmic Snake}

To build a realistic model to represent the Cosmic Snake, we distinguish two gas disks, one for the molecular component, and the second for the ionized gas component. Given that we are observing only the baryon-dominated central parts of a relatively massive galaxy, the contribution from the dark matter is negligible. Also the bulge component was neglected since it does not strongly affect the gas kinematics.

\begin{figure}[t]
\centering
\includegraphics[scale=0.58]{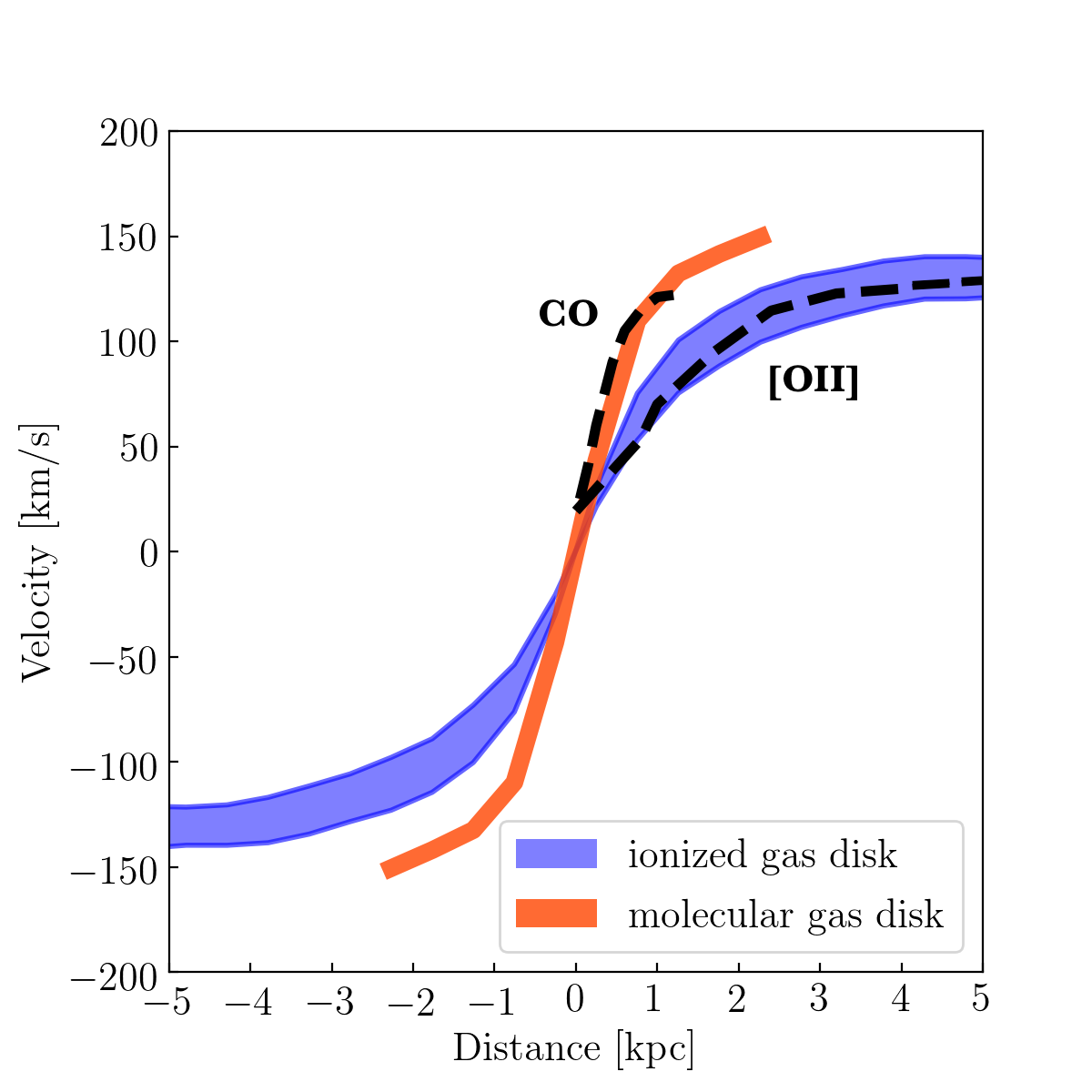}
\caption{Rotation curves obtained in the simulated mass distribution corresponding to the parameters listed in Table \ref{model_snake}, with two tracers: the molecular and ionized components. The results from the simulation for the molecular and ionized gas disks are shown in orange and in blue, respectively. The upper and lower boundaries of the ionized gas disk (taking the absolute values) represent a height of 1 kpc and 3 kpc, respectively. The dashed black lines indicate the observed CO and \oii \, rotation curves of the Cosmic Snake. }
\label{figure_model_snake}
\end{figure}

\begin{figure}[t]
\centering
\includegraphics[scale=0.58]{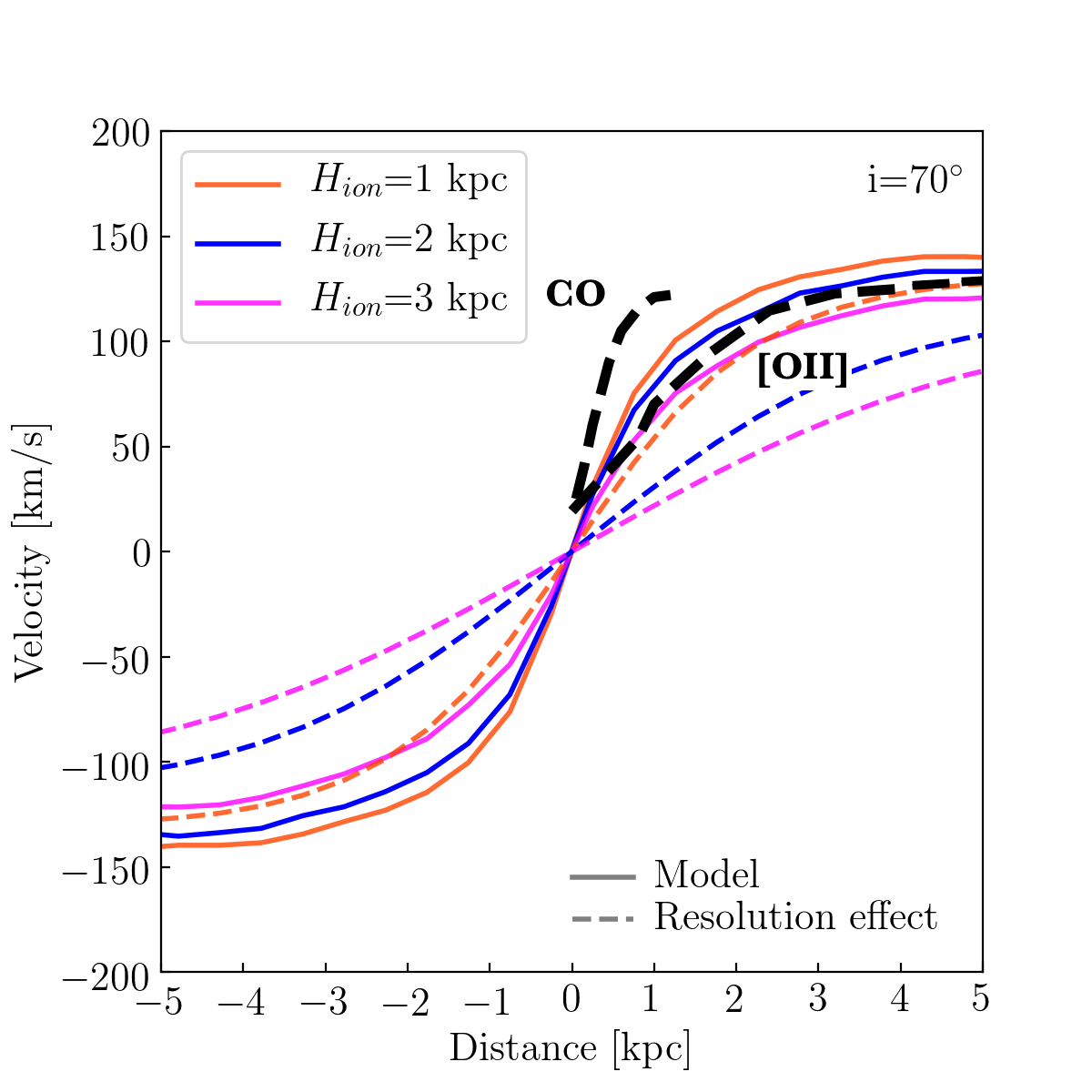}
\caption{Effect of the ionized gas disk height on the rotation curve due to the spatial resolution for the Cosmic Snake.  The coloured solid curves are the rotation curves simulated in the mass distribution corresponding to the parameters listed in Table \ref{model_snake} with an ionized disk height of 1 kpc (orange), 2 kpc (blue) and 3 kpc (pink). The dashed coloured curves represent the curves smoothed to the spatial resolution determined by Eq. \ref{resolution} with their corresponding disk height. The inclination has been fixed to $i=70^\circ$, which is the inclination of the Cosmic Snake. The dashed black lines indicate the observed CO and \oii \, rotation curves of the Cosmic Snake.}
\label{variePSF_2}
\end{figure}

For the stellar disk, the stellar mass is taken from the work of \citet{Cava2018}. We measure a stellar effective radius of $R_\star=3.8\pm0.7$ kpc from the radial mass profile, and  we adopt a stellar scale height typical of nearby disk galaxies with $H_\star\sim0.5$ kpc \citep[e.g.,][]{vanderKruit2011,Ma2017}.
From the CO luminosity profile, we measure the molecular effective radius to be $R_{mol}=0.75^{+0.26}_{-0.07}$ kpc and estimate the molecular height to be  $H_{mol}<450$ pc. We choose to fix the height to 100 pc since using a value between 50 pc and 450 pc does not strongly influence the results (see Fig. \ref{varie_rgas}). The mass of the molecular gas disk is determined from the gas fraction of $f_{gas}=0.20\pm0.03$ obtained in Dessauges-Zavadsky et al. (accepted). We estimate the ionized gas mass to be $(2.0\pm0.2)\times10^7 M_\odot$ from the absolute B magnitude following the relation of \citet{LopezSanchez2010}. This mass is very small in comparison to the total mass of the galaxy ($f_{ion}<<1\%$) and does not contribute to the potential of the galaxy. Our results are therefore not sensitive to this estimate. We derive an effective radius of $R_{ion}=6.1^{+1.4}_{-1.0}$ kpc for the ionized gas disk from the  radial \oii \, profile. We vary the ionized gas height between $1-3$ kpc, as it does not affect the potential of the galaxy.
Finally, the \citep{Toomre1964} stability parameter Q is directly measured using the relation

\begin{ceqn}
\begin{align}
\label{Eqtoomre}
Q =\sqrt{2}\frac{\sigma_0}{\upsilon_{rot} \, f_{gas}},
\end{align}
\end{ceqn}
where the instrinsic velocity dispersion is $\sigma_0\sim19.5 \pm6$ $\rm km \, s^{-1}$ (from this work), the rotation velocity corrected for the inclination is $\upsilon_{rot}=225\pm1$ $\rm km \, s^{-1}$ \citep{Patricio2018}, and the gas fraction is $f_{gas}=0.20\pm0.03$ (Dessauges-Zavadsky et al. accepted). We, therefore, get $ Q\sim0.6$ for both the molecular and ionized gas disks, since they show a similar velocity dispersion. We assume an inclination, $i$, of $70^\circ \pm 1^\circ$ for the molecular and ionized gas disk according to the kinematic model obtained by \citet{Patricio2018} for this galaxy. A difference in inclination between these two disks is not expected since the ionized gas results from the stars that form inside the molecular clouds. The parameters are listed in Table \ref{model_snake}.

Fig. \ref{figure_model_snake} shows the simulated molecular and ionized gas rotation curves together with the observed rotation curves from the CO and \oii \,  of the Cosmic Snake. The simulations are in very good agreement with the observed rotation curves. The rotation curves of the molecular and ionized gas are well reproduced by disks with different radii and heights. This means that the molecular gas disk of the Cosmic Snake is likely more concentrated in the center of the galaxy and has probably a thinner disk than the ionized gas disk. Galaxies with a thin molecular disk are also typically observed in the local Universe \citep[e.g.,][]{Glazebrook2013}. Some observations of local galaxies also reveal that the molecular gas is more concentrated in the center of several galaxies \citep[e.g.,][]{Wong2002,Leroy2008}.

This is in  agreement with the relation relating the radius and disk height to the rotation velocity and velocity dispersion of the disk \citep[e.g.][]{Genzel2011,White2017}: 

\begin{ceqn}
\begin{align}
\label{disk_thick}
\frac{\upsilon_{rot}}{\sigma_0} \approx \frac{R}{H},
\end{align}
\end{ceqn}
where $R$ and $H$ are the radius and height of the disk. In the Cosmic Snake, we obtain a similar intrinsic velocity dispersion and rotation velocity in the ionized and molecular gas disks, meaning that both disks have similar $\upsilon_{rot}/\sigma_0$ ratios and radius-to-height, $R/H$, ratios. This is consistent with smaller values for both the radius and height of the molecular gas disk compared to the ionized gas disk.

From this model, we can also test how the presence of an ionized gas thick disk degrades the spatial resolution along the line of sight in the Cosmic Snake and how it affects the shape of the rotation curve. We convolve the 2D ionized gas velocity maps with an ionized gas disk height, $H_{ion}$,  of 1 kpc, 2 kpc, and 3 kpc to get velocity maps at the spatial resolution determined by Eq. \ref{resolution} with their corresponding height and at the inclination of the Cosmic Snake ($i=70^\circ$). We can then extract the respective rotation curves. Fig. \ref{variePSF_2} shows the simulated ionized gas rotation curves for an ionized gas disk height of 1 kpc (orange), 2 kpc (blue) and 3 kpc (pink) together with the rotation curves  smoothed due to the lower spatial resolution. First, we see that an ionized gas height of $\sim1$ kpc fits relatively well the observations. We also find that the spatial resolution effect gets stronger for a thicker disk of $H_{ion}= 2$ kpc and $H_{ion}=3$ kpc. It means that the lower spatial resolution caused by the presence of a thick disk plays a major role in shaping the ionized gas rotation curve of the Cosmic Snake due to the high inclination of $i=70^\circ$ of this galaxy. As seen in Sect.\ref{sect_res_effect}, a thick disk affects even more the shape of the rotation curve, due to this effect, in the case of a more inclined disk.


\begin{table}[tb]
\caption[]{Parameters of the model for the Cosmic Snake}
\label{model_snake}
\centering                         
\begin{tabular}{l c }      
\hline\hline               
Parameters              &       Values                  \\ 
\hline                       
\noalign{\smallskip}
$M_\star [M_\odot$]     &    $ (4.0\pm0.5)\times 10^{10} $        \\
$R_\star$ [kpc]         &       $ 3.8 \pm 0.7      $           \\
$H_\star$ [kpc]         &         0.5                   \\
\noalign{\smallskip}
\hline    
\noalign{\smallskip}
$f_{mol} $              &      $  0.20 \pm 0.03$                  \\
$R_{mol}$ [kpc]         &      $  0.75^{+0.26}_{-0.07} $           \\  
$H_{mol}$ [kpc]         &         0.1                  \\
$Q_{mol}$               &         0.6                   \\
\noalign{\smallskip}
\hline    
\noalign{\smallskip}
$M_{ion} [M_\odot$]     &    $ (2.0\pm 0.2) \times10^7$       \\
$R_{ion}$ [kpc]         &     $   6.1^{+1.4}_{-1.0}        $       \\ 
$H_{ion}$ [kpc]         &         1.0 -- 3.0               \\
$Q_{ion}$               &         0.6                   \\
\noalign{\smallskip}
\hline    
\noalign{\smallskip}
$ i $ [$^\circ$]        &        $ 70 \pm1$                  \\

\noalign{\smallskip}
\hline    
\end{tabular}
\end{table}

\subsection{A521}

\label{sect_a521}

In contrast to the Cosmic Snake, the \oii \, and CO rotation curves are in good agreement in the galaxy A521 similar to the result of \citet{Ubler2018}.
However, when we undertake simulations at the spatial resolution observed in A521 ($\sim 2-3$ kpc), we obtain that the difference between the \oii \, and CO rotation curves observed in the Cosmic Snake at a spatial resolution of few hundred parsecs ($\sim500$ pc) is much less pronounced.The lack of spatial resolution can easily prevent to see a difference between two rotation curves. However, since the CO emission in A521 is also much more radially extended ($R_{mol} = 5.8\pm 1.0$ kpc) with respect to \oii \, ($R_{ion} = 6.5\pm 1.2$ kpc), one does not necessarily expect a molecular gas rotation curve as steep as for the Cosmic Snake according to Fig.~\ref{varie_rgas}. 

We also observe a very different intrinsic velocity dispersion for the molecular and ionized gas in this galaxy with values of $\rm 11\pm8 \,  km \, s^{-1}$ and $\rm 54\pm11 \,  km \, s^{-1}$, respectively. \citet{Ubler2018} found a similar value of $\rm \sim15-30 \,  km \, s^{-1}$ from both the molecular and ionized gas in their galaxy at $z\sim1.4$. Values of $\rm \sim10-20 \,  km \, s^{-1}$ and $\rm \sim20-40 \,  km \, s^{-1}$ for the  molecular and ionized gas, respectively, have been found by \citet{Cortese2017} at $z\sim0.2$, which are very similar to what is found for typical spiral galaxies at $z\sim0$ \citep[e.g.,][]{Epinat2010,Levy2018}. The velocity dispersion of the molecular gas of A521 is similar to galaxies in the local Universe. However, the velocity dispersion from the ionized gas of A521 is higher than galaxies in the local Universe, but this is also observed for high-redshift galaxies in many studies \citep[e.g.,][]{ForsterSchreiber2009,Wisnioski2015,Johnson2018,Girard2018a}. This could mean that the ionized gas disk is more turbulent than the molecular disk and could indicate the presence of a thick disk in this galaxy. Given also the high inclination of 72$^\circ$ in A521, a difference in thickness of the molecular and ionized gas disks should cause a difference between the two rotation curves according to Fig. \ref{variePSF_i} and Fig. \ref{variePSF_2}. It is therefore possible that the molecular gas has an intrinsically steeper rotation curve, but that the available spatial resolution does not resolve it.

\section{Conclusions}
\label{section6}

We have presented the molecular gas kinematics from ALMA CO(4-3) observations of two highly magnified galaxies at $z~\sim~1$ and compared it with the \oii \, kinematics obtained with MUSE/VLT \citep{Patricio2018}. These two galaxies, named here the Cosmic Snake and A521, are located behind the clusters MACSJ1206.2-0847 and Abell 0521.

We derive the observed rotation curves and velocity dispersion profiles in the image plane for both galaxies. For the Cosmic Snake, where we obtain a spatial resolution of only a few hundred parsecs on the major axis, we observe a difference between the molecular and ionized gas rotation curves in the inner part of the galaxy ($0-2$ kpc). The molecular gas curve is steeper than the ionized gas curve.
In A521, the molecular and ionized gas rotation curves agree, but the spatial resolution is only of a few kpc on the major axis.

Using simulations, we look at the effect of the effective radius and thickness of the gas disk on the observed rotation curve and find that a more extended and thicker disk smooths the curve.
We also find that the presence of a thick disk ($H_{gas}>1$ kpc), when the galaxy is strongly inclined ($i>70^\circ$), can smooth the rotation curve because it degrades the spatial resolution along the line of sight. 

We reproduce the observed rotation curves of the Cosmic Snake using a model that includes a stellar disk and two gas disks, with a molecular gas disk that is more massive and more radially and vertically concentrated than the ionized gas disk. The  model suggests a thick ionized gas disk with a scale height of $H_{ion}\sim1$ kpc. We measure an intrinsic velocity dispersion of $18.5\pm7$  $\rm km \, s^{-1}$   and $19.5\pm6$  $\rm km \, s^{-1}$ for the molecular and ionized gas of the Cosmic Snake. Similar values of the maximum rotation velocity and intrinsic velocity dispersion agree with a molecular gas disk that is more concentrated in the center of the galaxy and thinner than the ionized gas disk. For A521, we obtain values of $11\pm8$  $\rm km \, s^{-1}$   and $54\pm11$  $\rm km \, s^{-1}$ for the velocity dispersion, with a higher value for the ionized gas. This difference could indicate the presence of a thick and turbulent ionized gas disk.

Taken together, these two galaxies highlight the diversity in the kinematics observed at $z\sim1$. Their kinematics reflect the different spatial distribution of the molecular and ionized gas disks and confirm the presence of thick ionized gas disks at this epoch. These thick ionized gas disks are believed to be the result of the turbulence in the gas of young gas-rich galaxies \citep[e.g.,][]{ForsterSchreiber2009}.
A thinner and less radially extended molecular gas disk could also indicate that the formation of the molecular gas is limited to the central regions in some galaxies because of the low pressure or lack of metals outside the midplane and center of the galaxy \citep[e.g.,][]{Wong2002}.

\begin{acknowledgements}
This work was supported by the Swiss National Science Foundation. MG is grateful to the Fonds de recherche du Qu\'ebec - Nature et Technologies (FRQNT) for financial support. JR acknowledges support from the ERC Starting Grant 336736-CALENDS.
Based on observations made with the NASA/ESA Hubble Space Telescope, obtained from the Data Archive at the Space Telescope Science Institute, which is operated by the Association of Universities for Research in Astronomy, Inc., under NASA contract NAS 5-26555. These observations are associated with program \#15435.
     
\end{acknowledgements}

\bibliographystyle{aa} 
\bibliography{references.bib} 

\end{document}